\documentclass{Interspeech2024}
\usepackage{multirow}
\usepackage{xurl}
\usepackage{dirtytalk} 
\usepackage{cite}

\interspeechcameraready

\title{Gender Representation in TV and Radio: Automatic Information Extraction methods versus Manual Analyses}

\name[affiliation={1}]{David}{Doukhan}
\name[affiliation={2}]{Lena}{Dodson}
\name[affiliation={2}]{Manon}{Conan}
\name[affiliation={1}]{Valentin}{Pelloin}
\name[affiliation={2}]{Aurélien}{Clamouse}
\name[affiliation={2}]{Mélina}{Lepape}
\name[affiliation={2}]{Géraldine}{Van Hille}
\name[affiliation={3}]{Cécile}{Méadel}
\name[affiliation={4}]{Marlène}{Coulomb-Gully}

\address{
  $^1$INA, France
  $^2$ARCOM, France
  $^3$CARISM, France
  $^4$LERASS, France
  }
\email{\{ddoukhan, vpelloin\}@ina.fr, \{firstname.name\}@arcom.fr, cecile.meadel@u-paris2.fr, marlene.coulomb@univ-tlse2.fr}


\begin{document}

\maketitle

\begin{abstract}
    

This study investigates the relationship between automatic information extraction descriptors and manual analyses to describe gender representation disparities in TV and Radio. Automatic descriptors, including speech time, facial categorization and speech transcriptions are compared with channel reports on a vast 32,000-hour corpus of French broadcasts from 2023. Findings reveal systemic gender imbalances, with women underrepresented compared to men across all descriptors. Notably, manual channel reports show higher women’s presence than automatic estimates and references to women are lower than their speech time. Descriptors share common dynamics during high and low audiences, war coverage, or private versus public channels.
While women are more visible than audible in French TV, this trend is inverted in news with unseen journalists depicting male protagonists.
A statistical test shows 3 main effects influencing references to women: program category, channel and speaker gender.


    
    

\end{abstract}

\section{Introduction}









The media, akin to the family, school or courts, are depicted by philosopher Michel Foucault as \textit{power technologies}, contributing to the establishment of norms, while pretending to mirror society~\cite{foucault1994dits}.
Building on this concept, Teresa de Lauretis characterizes the media as \textit{gender technologies}, highlighting that gender representation constructs gender itself~\cite{de2007theorie}.
This construction, when women are underrepresented or depicted through gendered stereotypes, directly fuel their marginalization. Therefore, describing gender representation differences in the media  is necessary for improving social equality.


Quantitative studies are often viewed as authoritative arguments, necessary to obtain conviction and enable  political action against systemic gender stereotypes in media~\cite{porter1996trust, biscarrat2017soulevent}.
A 
key limitation of these approaches is their 
tendency to oversimplify
complex phenomena into ill-defined categories or numerical estimates.
This \textit{strategic essentialization}, as described by Spivak~\cite{spivak85}, deliberately reduces the complexity of gender identity and nuances into more measurable forms. Despite its flaws, this strategic approach is a pragmatic choice made to enable the description of gender representation biases in the media.

Gender monitoring studies face several practical challenges requiring to adopt the most suitable trade-offs between sample-size and analysis details, to produce the most significant knowledge using limited resources. 
Recent advances in machine learning have given rise to computational methods leveraging sample size issues associated with manual analyses
~\cite{doukhan2018describing,gay2023radio,mazieres2021computational,hong2021analysis,jurgens2022age,richard2022genderednews}. 
The exhaustiveness, along with the hype surrounding machine learning, contributes to the credibility and social impact of the resulting estimates.
However, if these indicators are obtained on different corpora and present disparate numerical values, less is known about the relations existing between them.

This study proposes employing various automated methods to estimate women's speech time, facial exposure, and quotation percentage in a large dataset of French broadcasts in 2023.
Automatic descriptors are compared to manual women presence reports provided by French channels to ARCOM: 
a French independent administrative agency in charge 
of promoting a fair representation of Women in TV and radio programs. 
The relationship between manual and automatic descriptors are investigated, considering factors such as program type, channel status (public/private), medium (TV/radio), peak time audience.


\section{Related Work}

Manual analyses generally produce a \textit{women presence rate} (\textbf{WPR}) metric, by counting  characters depicted in media programs.
Varying criteria are defined to state if a person should be included in the count, impacting studies sample size.
French GMMP TV and Radio news corpora have a relatively modest duration (36h), 
but requires to describe any appearing of evoked character, resulting in a coding density of 72 characters/hour of program~\cite{gmmp24, macharia2020global}.
ARCOM's diversity report restrict counts to speaking character, resulting in analyses based on 2672 programs, with 17 coded characters / program~\cite{arcom2023barometre}. 
ARCOMS's women representation report is much larger (41\,604 programs), but character's coding is reported by channels and is limited to characters depicted in studio conditions resulting in lower amounts of reported characters per program (3.9)~\cite{arcom2023femmes}. %
Manual studies going beyond WPR reporting are scarce and generally based on smaller amounts of data, reporting women speaking-time percentage and oral references to female characters~\cite{reiser2008}.

Several recent computational studies  proposed estimates based on women speaking time estimation ~\cite{doukhan2018describing, gay2023radio} or facial detection and categorization~\cite{mazieres2021computational,hong2021analysis,jurgens2022age}.
NLP methods were  also proposed for real-time monitoring of references to women in online press, based on gendered first name dictionaries and quotation detectors~\cite{richard2022genderednews}.
The few proposals mixing audio, visual or textual modalities report different conclusions.
\cite{geena} compared women speech-time to facial exposure in movies, concluding these two estimators produce equivalent average results but differs according to movie category.
\cite{doukhan2019estimer} compared speech-time, facial exposure and ASR-based references to women in raw TV streams, concluding women facial and lexical descriptors are higher than women speech-time.
\cite{gmmp24} compared speech time and visual exposure to manual WPR on GMMP French news corpus, reporting lower WPR than automatic estimates, together with more women speech than women faces.











\section{Method}

\subsection{ARCOM's Corpus}

ARCOM has been producing an annual public report since 2016, which includes quantitative and qualitative analyses, along with a list of recommendations intended for TV and radio channels to improve the representation of women in programs.
The quantitative aspects of this study rely on reports from French channels, as mandated by French law~\cite{loiarcom}.
Channels provide ARCOM with a list of the programs broadcasted during two \textit{neutral} months, excluding fictions (movies, cartoons).
ARCOM  communicates these months in advance of the monitored period to avoid major events that could affect women's representation, such as elections or sporting events. 
Programs are classified using a hierarchical taxonomy with 4 main categories: news, entertainments, magazines or documentaries, sport.
Each broadcast is associated with a report that includes the number of speaking characters appearing in studio conditions, categorized by gender (male/female) and by 5 roles: presenter, journalist, political guest, expert, and other.


The data considered for ARCOM's 2023 report is based on the analysis of May and October 2023, including 26 TV 
and 15 radio channels~\cite{arcom24}. A total of 29\,707 programs and 135\,385 individuals were reported by channels to ARCOM.
Each reported broadcast is associated to start and end time and date, which can  be either theoretical (broadcast schedule) or obtained from real-time broadcast monitoring.
Peak audience time slots are defined as 6AM-9AM for radio and 6PM-11PM for TV.
Accurate time codes corresponding to commercial breaks (in-show advertisements) are available for a subset of 19 TV channels (excluding radio and continuous TV news) and obtained from Médiamétrie's commercial dataset.\footnote{\url{https://www.mediametrie.fr/}} 
The corpus has a global duration of 31\,787 hours (about 1324 days), with an average 4.3 reported persons per broadcast hour.






\subsection{Speaker Gender Segmentation}

Speaker gender and Voice Activity Detection (VAD) were estimated using \texttt{inaSpeechSegmenter v0.7.7}~\cite{doukhan2018open}.
This software uses a definition of speech activity (\texttt{smn} VAD module) restricted to spoken speech and  does not predict singing voice gender, which is classified as music and excluded from speech time analyses.
This software based on relatively shallow CNNs required an average 30 seconds to process 1 hour of audiovisual data (267 hours in total) using old generation GPU (GeForce Titan X
). A benchmark realized using \texttt{InaGVAD}, a challenging TV and Radio corpus representative of French audiovisual diversity, showed its superior performance over 6 state-of-the-art VAD systems (Accuracy = 93\%) and 3 speaker gender segmentation systems (WSTP global estimation error = -0.1\%)~\cite{inagvad}.

\subsection{Augmented speech transcriptions}

Automatic speech transcriptions were obtained using Whisper's \texttt{large-v3} multilingual model~\cite{radford2023robust}.
WhisperX framework was used to speed-up transcription  using \texttt{faster-whisper} backend, and to obtain more accurate 
utterance timestamps using a \text{wav2vec2} alignment procedure suited to French~\cite{bain23_interspeech}.
This process required about 4 minutes per hour of audiovisual data (89 days for the whole corpus) using GeForce RTX 2080 Ti. 
The most common hallucination patterns were excluded using regex \verb1^\W*sous-titr(age|es par)1. 
Automatic transcriptions were aligned with \texttt{inaSpeechSegmenter}'s output (see previous section) in order to associate  each utterance to its corresponding duration of externally detected male and female speech. 
This choice allowed the dedicated filtering of irrelevant utterances depending on the phenomena being described and a better management of Whisper’s known characteristics: word deletion in speech overlapping situations involving several speakers, singing voice transcription and hallucinations.
By the end of this process, 22M utterances and 271M words were obtained.

\subsection{References to Men and Women}

The number of oral references to men and women was approximated using a lexicometric approach based on the counting of gendered first names.
Names were extracted from automatic speech transcripts using an open database\footnote{\url{https://www.insee.fr/fr/statistiques/2540004}}  maintained by INSEE (the French National Institute of Statistics and Economic Studies), which contains detailed information about first names given to children born in France from 1900 to 2021
, including birth year and sex assigned at birth.
This data enabled the estimation of first name frequency in the French population, as well as gender attribution percentage of gender-neutral first names (ie: the name \say{Claude} was attributed to 468,462 individuals, including 88\% male and 12\% female).
Similarly to the Gendered News project~\cite{richard2022genderednews}, we chose to map each detected first name to its gender attribution percentage, in order to include gender-neutral names in the estimation of the percentage of references to women.
Capitalized symbols and punctuation marks predicted by Whisper were used to filter words found in the first name lexicon.
We exclude from counts lowercase words (common nouns), 
fully capitalized word (acronyms),
and consecutive first names separated by whitespace characters (e.g. \say{Gazi Mustafa Kemal} is associated with a single count corresponding to \say{Gazi}).
For each name category (male, female and gender-neutral), a manual inspection was carried out on the top 150 names with the highest ratio of occurrences in ARCOM's corpus divided by occurrences in the French population. This selection included various phenomena, including international personalities having uncommon names in France (Bachar, Benyamin, Vladimir), homonyms (Niagara, Nirvana, Amen, Tennessee, Eureka...), last names detected as uncommon first names (Wagner, Ndiaye, ...). The frequency of these difficult examples was about 2\% for each first name categories.
We eventually chose to retain all detected first names, considering that the impact of these outliers on our resulting estimators would be minimal and that this choice would allow us to obtain estimates in fully automatic conditions. By the end of this process, 3.2 million first names were extracted.

\subsection{Face Detection and Categorization}

This study introduces \texttt{inaFaceAnalyzer}, a novel open-source facial analysis framework designed for analyzing large amounts of audiovisual data.\footnote{\url{https://github.com/ina-foss/inaFaceAnalyzer}} 
Face detection was performed using OpenCV CNN back-end, with a detection confidence threshold of 0.65 and a facial bounding box scaling factor of 1.1 suited to optimal gender classification performances.
Detected faces were classified using a Resnet50 DNN~\cite{he2016deep} trained using FairFace~\cite{karkkainenfairface}, a balanced gender and racial face-in-the-wild dataset.
Resulting model achieved a 95.3\% face gender classification accuracy 
and was optimized to minimize 
gender, age and racial biases defined as
classification error differences between men and women (+0.8) , young and 50+ old adults (-1.5), white and non-white individuals (-0.2) based on FairFace annotations.
Cross-corpora evaluations detailed on project's home-page report superior accuracy and lower classification biases over 2 open-source baselines.
Face gender predictions were obtained using a frame analysis rate of 1 image per second.
Faces were filtered to keep only those associated to a minimal height of 10\% of image frame, for technical (smaller faces are associated to lower classification performance) and practical reasons (larger faces appearing on screen are the most relevant for describing representation).
Facial analyses were distributed in a heterogeneous CPU cluster, using between 16 and 32 cpu per process.
Average inference processing time was close to 3 minutes per hour of TV program, requiring 773 hours of computations (32 days) for processing 15,471 hours of TV programs.
14.5M women and 25.3M men faces were obtained through this process, corresponding to a ratio of 0.71 faces per seconds.

\section{Results}

Women Representation Estimates (\textbf{WRE}) are presented using four acronyms. \textbf{WPR} (Women Presence Rate) corresponding to manual presence rates reported by channels to ARCOM. \textbf{WSR} (Women Speech Rate) defined as the percentage of speech uttered by women. \textbf{WQR} (Women Quotation Rate), referred to as the percentage of women first names. \textbf{WFR} (Women Face Rate) defined as the percentage of women faces detected. All these metrics are defined such that percentages of both genders add up to 100: estimates of 50\% imply that a descriptor is fairly represented across gender categories.


\subsection{Women representation in TV commercial breaks}

Commercial breaks represented an average of 9.5\% of the duration of ARCOM corpus subset, with lower percentages observed in public channels (3.3\%).
Highest advertisement durations were found in private channels (11.5\%), reaching up to 20.5\%. 
The automatic WRE associated with isolated advertisement excerpts versus TV programs without advertisements are detailed in table~\ref{tab:pubnopub}. This table shows a greater  representation of women in advertising slots across all indicators, with the largest effects observed for women's speech (+9.7\%), women's faces (+7.7\%) and lower effects for women's first name percentages (+2.4\%).
Beyond noting the increased presence of women in advertisements compared to the rest of the programs, this result highlights one limitation of our methodology: 
the lack of in-show advertisements time codes for all channels.
Additionally, the quantity of commercial breaks varies per channel, making it difficult to model advertisement effects in channel where this information is unknown.
We chose to exclude in-show advertisements, when this data was available, from the results presented in the next sections of this study.
This trade-off may lead to an overestimation of women's presence in continuous TV news channels and radio, particularly in private channels.


\begin{table}
  \caption{Automatic Women presence percentage in TV programs and in-show advertisements}
  \label{tab:pubnopub}
  \centering
  \begin{tabular}{lcc}
    \toprule
    \textbf{Descriptor}      & \textbf{TV program} & \textbf{Advertisements}               \\
    \midrule
    Speech \%                  & 33.6  &   43.3    \\
    Face \%                  & 40.2       &   47.9  \\
    First Names \%       & 33.8      &  36.2  \\
    \bottomrule
  \end{tabular}
\end{table}

\subsection{Women representation descriptor dynamics}

Tables \ref{tab:audience} and \ref{tab:privpub} present WRE variations according to medium (TV/radio), channel status (public/private) and broadcast timing (low or high audience timeslot).
Firstly, it's worth noting that in both manual and automatic configurations, WRE are below 50\%, which reminds that French TV and radio still quantitatively under-represent women in favor of men.
Furthermore, in each configuration, manual WPR is higher than any other automatic descriptor.
Similarly, when available for TV, estimated WFR is always greater than WSR, suggesting that women are more visible than audible in French TV.
Additionally, WQR is consistently lower than WSR, indicating that TV and radio speakers predominantly depict male characters.

According to all descriptors, WRE is higher on radio during high audience time-slots, while an opposite trend is observed for TV with fewer women appearing during high audience programs.
Similarly and for all indicators, highest amounts of women are depicted in public than in private channels.
These findings highlight the common dynamics shared by these four indicators, each with different natures and meanings.

\begin{table}
  \caption{Women (\%) in high and low audience programs}
  \label{tab:audience}
  \centering
  \begin{tabular}{lccccc}
    \toprule
    \textbf{Media} & \textbf{Audience} & \textbf{Manual} & \textbf{Speech} & \textbf{Name} & \textbf{Face}   \\
    \midrule
    Radio & low & 41.3 & 33.4 & 29.3\\
    Radio & high & 44.0 &  34.9 & 33.0\\
    TV & low & 41.5 & 36.0 & 32.9 & 37.2 \\
    TV & high & 37.5 & 31.7 & 30.6 & 33.9\\
    \bottomrule
  \end{tabular}
\end{table}

\begin{table}
  \caption{Women (\%) in public and private channels} 
  \label{tab:privpub}
  \centering
  \begin{tabular}{lccccc}
    \toprule
    \textbf{Media} & \textbf{chan. status} & \textbf{Manual} & \textbf{Speech} & \textbf{Name} & \textbf{Face}   \\
    \midrule
    Radio & private & 36.0 & 28.3 & 28.2 & \\
    Radio & public & 44.8 & 37.9 & 31.7 & \\
    TV &  private & 39.7 & 32.0 & 31.3 & 34.6 \\
    TV & public & 45.0 & 42.1 & 34.9 & 40.3 \\
    \bottomrule
  \end{tabular}
\end{table}

\subsection{WRE Distributions by TV program categories} 

A detailed WRE analysis according to TV program categories reveals distinct distributions
 detailed in table \ref{tab:tvprogtype}.
Lowest WRE were found in the sport category, with TV automatic descriptors being half as low as reported manual indicator (10.9-12.0\% vs 21.5\%).
TV entertainments WFR was found to be the only automatic indicator reporting larger women representation than its corresponding manual report.
WFR in TV entertainments was also 10.7 points above  WSR: the largest difference found between the visual and speech modalities. This important difference can be partly explained by a large amount of musical performances found in TV entertainments: associated to singing speech which is discarded from analyses, while the corresponding faces appearing on screen (singer, dancers, audience...) are taken into account.
While WFR is generally higher than WSR (see last section), TV news were associated to an opposite tendency.
Manual analyses were carried on a subset of 200 short TV news programs and revealed this phenomena was due to large amounts of women journalists, invisible on-screen, describing news in which men are the main protagonists and are depicted on screen.
This trend is also observed for WQR, which is also lower in TV news than in Entertainments or Magazines.

\begin{table}
  \caption{TV Women Presence percentage per program type} 
  \label{tab:tvprogtype}
  \centering
  \begin{tabular}{lcccc}
    \toprule
     \textbf{Program type} & \textbf{Manual} & \textbf{Speech} & \textbf{Name} & \textbf{Face}   \\
    \midrule
Entertainment & 44.0 & 36.0 & 43.1 & 46.7\\
News & 42.6 & 37.1 & 30.4 & 33.1\\
Magazine/Documentary & 46.7 & 38.3 & 39.1 & 45.0\\
Sport & 21.5 & 11.4 & 10.9 & 12.0\\
    \bottomrule
  \end{tabular}
\end{table}


\subsection{Armed conflict effects}

Women representation in news is partly influenced by the prevailing national and international events.
Although the two months selected prior to ARCOM's monitoring campaign were supposed to be \textit{neutral} (May and October 2023), 
the armed conflict involving Israel and Hamas gained significant prominence in French news broadcasts starting from October 7th.
The media coverage of this conflict resulted in lower women presence in news programs according to the majority of indicators detailed in table \ref{tab:hamas}. 
This can be partly attributed to the underrepresentation of women among 
Israeli and Hamas representatives, along with a known tendency in French media to solicit men (journalists, experts) to discuss armed conflicts. 

\begin{table}
  \caption{Women percentage difference in news programs after the beginning of Israel-Hamas conflict (October 7th, 2023)} 
  \label{tab:hamas}
  \centering
  \begin{tabular}{lccccc}
    \toprule
    \textbf{Media} & \textbf{chan. status} & \textbf{Manual} & \textbf{Speech} & \textbf{Name} & \textbf{Face}   \\
    \midrule
Radio & Private &  -2.4 & -0.2 & -3.3 & \\
Radio & Public & -5.9 & -2.2 & -1.3 & \\
Tv & Private & -1.0 & +1.7 &  -5.3 & -0.8\\
Tv &  Public & 0.0 & -2.6 & -1.6 & -3.3\\
    \bottomrule
  \end{tabular}
\end{table}

\subsection{Effects influencing References to Women Characters}

Effects influencing WQR were modelled from a population of 2.3M isolated utterances containing at least one first name, a minimum voice activity ratio of 50\% (excluding singing voice) and a non-ambiguous speaker gender estimation with women speech ratio below 20\% (mostly male) or above 80\% (mostly female).
Ordinal Least Square regression models (\texttt{statsmodels} package~\cite{seabold2010statsmodels}) were used to predict utterances first names gender attribution percentage mean based on 7 parameters: media (tv/radio), channel, status (public/private), program type, broadcast audience (low/high), armed conflict (before/after) and speaker gender.
Effect size was estimated from corresponding ANOVA tables using eta-squared ($\eta^2$)~\cite{levine2002eta}.
Statistically significant p-scores ($PR(>F)	< 0.05$) were obtained for each isolated parameters, at the exception of the public/private channel status. Small effects ($\eta^2>=0.01$) were observed for program category ($\eta^2 = 0.034$), channel ($\eta^2=0.015$) and speaker gender ($\eta^2=0.011$).
Female speaker pronounced 36.9\% of women's first names,  while male speakers uttered only 28.2\% women's first names.
Detailed estimates presented in table \ref{tab:namegender} show that highest WQR are uttered by female speakers during TV and Radio low-audience time-slots, with percentage differences of about 10 points compared to male speakers.
While WQR was equivalent for male and female speakers in radio-high audience programs, an 8 point difference was observed on TV.

\begin{table}
  \caption{Percentage of women first name uttered depending on speaker gender, medium and audience time-slot} 
  \label{tab:namegender}
  \centering
  \begin{tabular}{lccc}
    \toprule
    \multirow{2}[3]{*}{\textbf{Media}} & \multirow{2}[3]{*}{\textbf{Audience}} & \multicolumn{2}{c}{\textbf{Women first name \%}}\\ \cmidrule{3-4}
     & & female speaker & male speaker   \\
    \midrule
Radio & Low &  36.1 & 26.3\\
Radio & High & 32.6 & 32.8\\
Tv & Low & 39.0 & 29.2\\
Tv & High & 34.2 & 28.2\\
    \bottomrule
  \end{tabular}
\end{table}





\section{Discussion}


In this study, we illustrate the complex relationship between automatic descriptors of women's representation and manual channel reports on a large 32k hours TV and radio corpus.
While these descriptors share common dynamics, WPR obtained through manual channel reports were predominantly higher than that of automatic descriptors. 
This result is opposed to~\cite{gmmp24}, reporting opposite trends in the smaller French GMMP corpus (36h) using a more exhaustive methodology for counting women.
This finding suggests that the methodology used by the channels, despite being used on massive sample sizes, may overestimate women representation in French TV and Radio.
Manual inspection found significant disparities between manual and automatic estimations, particularly in programs with minimal studio time, reflecting differences between manual reports (studio-limited) and automatic analyses (entire programs).


While we managed to exclude in-show advertisements for 19 out of 41 channels 
, we also observed higher WRE in advertisements than in programs.
This finding illustrates possible biases in our proposal, which may result in WRE overestimation for channels broadcasting the largest amounts of advertisements: more specifically private radio and private continuous news TV channels - already associated with the lowest WRE.

Similarly to \cite{doukhan2019estimer}, we observed larger average WFR than WSR in TV programs.
However, analyses based on program categories show an inverted tendency in TV news program, consistent with results reported on the smaller GMMP  corpus~\cite{gmmp24}.

Among the largest statistically significant effects influencing Women quotation percentage, we show speaker gender resulted in large WQR differences, with female speakers pronouncing 
women's first names 
more often
(36.9\%)  than male speakers (28.2\%).
While qualitative studies reported homosociability effects~\cite{biscarrat2023ordre, doi:10.1177/10776990211042595}, further research  is needed to better isolate these effects from gendered distribution of  topics~\cite{pelloininterspeech24}.  


We should acknowledge our proposal contains limitations requiring additional research efforts.
While we used first names as a proxy for estimating the percentage of oral references, several studies reported women more often addressed by their first name than men, suggesting this descriptor may underestimate oral references to men~\cite{files2017speaker}.
Large differences between speech and face indicators in entertainment, containing large amounts of music and excluding singing voice from WSR, highlight the need for speaker gender prediction systems suited to the analysis of singing voices: a task known to be harder~\cite{kong2023strada}.
Lastly, our methodological framework required categorizing individuals into stereotypical binary genders.
It is worth noting that speech cues conveying elements of gender identity are contextually bound~\cite{holmes1998signalling,kachel2024speakers} 
and that the representation of openly non-binary individuals in French TV and radio is limited, 
making
it difficult to build systems allowing to predict 
gender identities based on purely acoustic, visual cues or lexical cues.




\section{Acknowledgements}
This work has been partially funded by 
     the French National Research Agency (project Gender Equality Monitor - ANR-19-CE38-0012).
Women representation channel reports per program are not publicly available and were obtained 
with the support of ARCOM.
The corresponding 32K hours audiovisual materials were obtained with the support of 
the French National Audiovisual Institute (INA) 
: a public institution in charge of the digitalization, preservation, distribution and dissemination of the French audiovisual heritage.
Due to obvious copyright concerns, these materials are not openly accessible.
Finally, authors would like to thank Albert Rilliard for its valuable help in conducting statistical tests.

\bibliographystyle{IEEEtran}
\bibliography{mybib}

\end{document}